%% file: Vasak_ARXIV.tex
\documentclass[aps,prl,preprint,groupedaddress]{revtex4-2}

\usepackage{amsmath}
\usepackage{tensor}
\usepackage{color}
\usepackage{verbatim}
\usepackage{nicefrac}

\newcommand{\hoc}{\tilde{q}}
\newcommand{\HCd}{\mathcal{H}}
\newcommand{\LCd}{\mathcal{L}}
\newcommand{\dd}{\mathrm{d}}
\newcommand{\pfrac}[2]{\frac{\partial{#1}}{\partial{#2}}}
\newcommand{\onehalf}{{\textstyle\frac{1}{2}}}

\newcommand{\quarter}{{\textstyle\frac{1}{4}}}

\newcommand{\threehalf}{{\textstyle\frac{3}{2}}}

\usepackage{mathtools}
\usepackage{amsmath}
\usepackage{palatino}

\begin{document}


\title{On the cosmological constant in the deformed Einstein-Cartan gauge gravity in De Donder-Weyl Hamiltonian formulation}

\author{D.~Vasak\thanks{vasak@fias.uni-frankfurt.de}}
\affiliation{Frankfurt Institute for Advanced Studies (FIAS), 
Ruth-Moufang-Strasse~1, 60438 Frankfurt am Main, Germany}

\author{J.~Struckmeier}
\affiliation{Frankfurt Institute for Advanced Studies (FIAS), 
Ruth-Moufang-Strasse~1, 60438 Frankfurt am Main, Germany}
\affiliation{Goethe Universit\"at, Max-von-Laue-Strasse~1, 60438~Frankfurt am 
Main, Germany}

\author{J.~Kirsch}
\affiliation{Frankfurt Institute for Advanced Studies (FIAS), 
Ruth-Moufang-Strasse~1, 60438 Frankfurt am Main, Germany}

\author{H.~St\"ocker}
\affiliation{Frankfurt Institute for Advanced Studies (FIAS), 
Ruth-Moufang-Strasse~1, 60438 Frankfurt am Main, Germany}
\affiliation{Goethe Universit\"at, Max-von-Laue-Strasse~1, 60438~Frankfurt am 
Main, Germany}
\affiliation{GSI Helmholtzzentrum f\"ur Schwerionenforschung GmbH, 
Planckstrasse~1, 64291 Darmstadt, Germany}
\date{\today}

\begin{abstract}
A modification of the Einstein-Hilbert theory, the Covariant Canonical Gauge Gravity (CCGG), leads to a cosmological constant
that represents the energy of the space-time continuum when deformed from its (A)dS ground state to a flat geometry.
CCGG is based on the canonical transformation theory in the De Donder-Weyl (DW) Hamiltonian formulation.
That framework modifies the Einstein-Hilbert Lagrangian of the free gravitational field by a quadratic Riemann-Cartan concomitant.
The theory predicts a total energy-momentum of the system of space-time and matter to vanish, in line with the conjecture of a ``Zero-Energy-Universe'' going back to Lorentz (1916) and Levi-Civita (1917).
Consequently a flat geometry can only exist in presence of matter where the bulk vacuum energy of matter, regardless of its value, is eliminated by the vacuum energy of space-time.
The observed cosmological constant $\Lambda_{\mathrm{obs}}$ is found to be merely a small correction
attributable to deviations from a flat geometry and effects of complex dynamical geometry of space-time, namely torsion and possibly also vacuum fluctuations.
That quadratic extension of General Relativity, anticipated already in 1918 by Einstein, 
thus provides a significant and natural contribution to resolving the
``cosmological constant problem''.
\end{abstract}


\maketitle


\section{Introduction}
The assumption that Einstein's cosmological constant represents vacuum energy 
fluctuations has caused what is called
the ``cosmological constant problem''~\citep{weinberg89,Carroll:2000fy}, or the worst theoretical estimate in the history of science.
The reason is that the calculated value of the field-theoretical vacuum energy of matter differs
from that deduced from astronomical observations by the huge factor of $\sim 10^{120}$.
In this paper we discuss a modification of the Einstein-Hilbert theory, based on a rigorous mathematical formalism, that provides a substantial contribution to resolving this 
{``problem''} within the realm of semi-classical gauge field theory of gravity.

\section{The Covariant Canonical Gauge theory of Gravity}
The mathematical framework underlying CCGG is the covariant canonical transformation theory, a framework well known from classical Hamiltonian mechanics, extended to the realm of relativistic field theories.
The underlying covariant, field-theoretical version of the canonical transformation theory in the De Donder-Weyl Hamiltonian formalism~\citep{dedonder30, struckmeier08} provides a stringent guidance for working out a gauge theory of gravity.
This means promoting a \emph{global}, i.e.\ Lorentz-invariant action of matter fields in a static space-time background, to a \emph{local}, i.e.\ Lorentz and diffeomorphism-invariant description in a dynamical space-time, thereby unambiguously fixing the \emph{coupling} between gravitational and matter fields.

In the CCGG framework, the non-degenerate\footnote{This means the regular Legendre transformation exists.} ``free'' (vacuum) gravity and matter Lagrangians are the initial input, in conjunction with the physical postulates of diffeomorphism invariance (Einstein's Principle of General Relativity \citep{einstein55}), and the Equivalence Principle, hence the existence of local inertial systems at any point of space-time.
The covariant canonical transformation formalism then yields the coupling terms of matter and gravitational fields that render the total system diffeomorphism invariant~\citep{struckmeier17a,struckmeier21a}.
The components of the gauge field are thereby identified with the connection coefficients.

Similar to all gauge theories, the dynamics of the free gauge field---which means here the dynamics of the gravitational field in source-free regions---is an independent input deduced separately on the basis of physical reasoning and subsequent experimental confirmation.
However, in contrast to other field theories, the current observational basis does not unambiguously determine the Lagrangian of the free gravitational field.
Beyond the Hilbert Lagrangian also formulations with various quadratic contractions of the Riemann or Riemann-Cartan tensor admit the Schwarzschild-de~Sitter and even the Kerr-de~Sitter metric as the solution of the pertaining field equations~\citep{stephenson58}.
Consequently, a combination of Einstein's linear ansatz with the ``Kretsch\-mann Lagrangian'', the latter consisting of the complete contraction of two Riemann tensors, is  {equally} a valid description of the dynamics of the free gravitational field even if torsion is admitted. Such a squared Riemann tensor invariant in the Lagrangian was even anticipated by Einstein already hundred years ago and suggested in his letter to Weyl~\citep{einstein18}.
And a free gravity Lagrangian with some quadratic concomitant of the Riemann tensor is in fact necessary~\citep{Benisty:2018ufz} to ensure the existence of a corresponding covariant DW Hamiltonian~\citep{dedonder30} by means of a Legendre transformation.

\section{The cosmological constant}
Here we review the relevant features of CCGG and show how a term with the 
properties of the cosmological constant \emph{emerges}  as a combination of two 
coupling constants of the theory.\footnote{The contribution of geometry to dark 
energy in CCGG has been discussed in~\cite{Vasak:2018gqn, vasak20, benisty21a}, see 
also~\cite{chen10} for a related ansatz.}  
This geometrical constant eliminates the bulk of vacuum energy of matter.

The properties of the theory and empirical insights combine to the following reasoning:
\begin{enumerate}
\item In CCGG with conventions as in~\cite{misner73}, the combined action of matter fields that interact with gravitational fields is~\citep{struckmeier21a}:
\begin{eqnarray}
S_0 &\!=\!\!&
\int_{V} \!\! \tilde{\LCd}_{\mathrm{tot}}\,\dd^4x
\equiv
\int_{V} \!\! \left(\tilde{\LCd}_{\mathrm{Gr}}+\tilde{\LCd}_{\mathrm{matter}}\right)\dd^4x \\
&\!=\!\!&
\int_{V} \!\!  \left(\tilde{k}\indices{_i^{\mu\nu}}\,S\indices{^i_{\mu\nu}}
+\onehalf \, \hoc\indices{_i^{j\mu\nu}}\,R\indices{^i_{j\mu\nu}}-\tilde{\HCd}_{\mathrm{Gr}}+\tilde{\LCd}_{\mathrm{matter}}\right)\dd^4x \, .\nonumber
\end{eqnarray}
The integrand $\tilde{\LCd}_{\mathrm{tot}}$ is the total Lagrangian consisting of the Lagrangians for the dynamical space-time coupled to matter, with the gravity Lagrangian expressed here as a Legendre transform of the corresponding DW Hamiltonian density $\tilde{\HCd}_{\mathrm{Gr}}$ of free gravity.
The Greek (Latin) letters denote holonomic (anholonomic) indices, respectively.
The former are lifted or contracted with the general metric $g_{\mu\nu}$  on the base manifold, the latter with the Minkowski metric $\eta_{ij} = \mathrm{diag} (1,-1,-1,-1)$ on the tangent space.
The expressions displayed are the dynamical fields of space-time -- torsion $S\indices{^i_{\mu\nu}}$ and Riemann-Cartan curvature tensor $R\indices{^i_{j\mu\nu}}$ -- expressed in terms of the vierbein field, $e\indices{^i_\mu}$, and the spin connection coefficients (the gauge field), $\omega\indices{^i_{j\nu}}$:
\begin{equation}
S\indices{^i_{\mu\nu}} = \onehalf \left(
\pfrac{e\indices{^i_\mu}}{x^\nu} - \pfrac{e\indices{^i_\nu}}{x^\mu}
+ \omega\indices{^i_{j\nu}} \, e\indices{^j_\mu} - \omega\indices{^i_{j\mu}} \, e\indices{^j_\nu} \right) \, ,
\end{equation}
\begin{equation}
R\indices{^i_{j\mu\nu}} = \pfrac{\omega\indices{^i_{j\nu}}}{x^\mu} - \pfrac{\omega\indices{^i_{j\mu}}}{x^\nu} + \omega\indices{^i_{n\mu}}\,\omega\indices{^n_{j\nu}} - \omega\indices{^i_{n\nu}}\,\omega\indices{^n_{j\mu}} \, .
\end{equation}

\noindent
The metric is given by the contraction of the vierbein fields w.r.t. the anholonomous indices, 
$g_{\mu\nu} \equiv \eta_{ij}\,e\indices{^i_\mu}\,e\indices{^j_\nu}$,  and we use 
\begin{equation}
\varepsilon:= \det \,e\indices{^i_\mu} \equiv \sqrt{-g_{\mu\nu}}
\, , 
\end{equation}
the determinant of the vierbein, for the invariant volume element and for defining relative tensors.
In the DW Hamiltonian formulation the ``velocity'' fields are ``traded'' via the Legendre transformation for the respective (conjugate) momentum tensor densities,
\setlength\abovedisplayskip{0pt}
  \setlength\belowdisplayskip{0pt}
\begin{equation}
 \tilde{k}\indices{_i^{\mu\nu}} \equiv k\indices{_i^{\mu\nu}}\varepsilon := \pfrac{\tilde{\LCd}_{\mathrm{tot}}}{e\indices{^i_{\mu,\nu}}} \, , 
 \end{equation}
\begin{equation}
\hoc\indices{_i^{j\alpha\beta}}\equiv q\indices{_i^{j\alpha\beta}}\varepsilon := \pfrac{\tilde{\LCd}_{\mathrm{tot}}}{\omega\indices{^i_{j\alpha,\beta}}} \, .
 \end{equation}
 with comma denoting partial derivative.\footnote{The formulation of Ref. \cite{struckmeier21a} is by construction metric compatible as the affine connection is a dependent field and can be designed to ensure metric compatibility.
In the metric formulation \citep{struckmeier17a} the affine connection is the independent field rather than the spin connection, and the momentum field $\tilde{k}\indices{_{\alpha\nu\mu}}$ is the conjugate of non-metricity. Which formulation is used is irrelevant for the key message of this paper, though.
}
\item By the necessity of the Legendre transformation between the Lagrangian and the DW Hamiltonian densities to exist, both must be non-degenerate.
The free gravity DW Hamiltonian must then include at least  the full quadratic tensor concomitant of the conjugate momenta~\citep{Benisty:2018ufz}.
Similarly to the free matter Hamiltonians that  {serve as} the key input to any gauge theory of gravitation,
also the free gravity Hamiltonian must be known in advance.
The usual way to obtain this DW Hamiltonian is to postulate it based on analogies with other field theories, and to experimentally confirm the solutions of the emerging field equations thereafter.
A reasonable choice for postulating $\tilde{\HCd}_{\mathrm{Gr}}(\tilde{q},\tilde{k},e)$ is 
the quadratic-linear ansatz~\citep{struckmeier21a}
\begin{align}
\tilde{\HCd}_{\mathrm{Gr}}&=\frac{1}{4g_{1}\varepsilon}\tilde{q}\indices{_{i}^{j\alpha\beta}}
\tilde{q}\indices{_{j}^{i\xi\lambda}}\,g_{\alpha\xi}\,g_{\beta\lambda} -
g_{2}\,\tilde{q}\indices{_{i}^{j\alpha\beta}}\,e\indices{^i_\alpha}\,e\indices{^n_\beta}\,\eta_{nj} \nonumber\\
&\quad+\frac{ {1}}{4g_3\varepsilon}\,\tilde{k}\indices{_{i}^{\alpha\beta}}
\tilde{k}\indices{_{j}^{\xi\lambda}}\,g_{\alpha\xi}\,g_{\beta\lambda}\,\eta^{ij},
\label{eq:ham-free-grav}
\end{align}
where $g_1$, $g_2$, and $g_3$ are \emph{fundamental} coupling constants which must be adapted to observations. Notice that no (cosmological) constant term is included here.
\item The dynamics of the system is given by the set of canonical equations arising by the variation of the action integral.
With the Hamiltonian \eqref{eq:ham-free-grav} the variation w.r.t.\ the momentum field $\tilde{q}\indices{_{i}^{j\alpha\beta}}$ (conjugate to the connection) leads to
\begin{equation}
q\indices{_{i}^{j\alpha\beta}} = g_1\left(R\indices{_{i}^{j\alpha\beta}} - \bar{R}\indices{_{i}^{j\alpha\beta}}\right),
\end{equation}
where
\begin{equation} \label{def:desitterR}
\bar{R}\indices{^{i}_{j\mu\nu}} := g_2 \left(
e\indices{^i_\mu}\,e\indices{^k_\nu} - e\indices{^i_\nu}\,e\indices{^k_\mu}
\right) \eta_{kj}
\end{equation}
 is the Riemann curvature tensor of the maximally symmetric space-time, i.e.~de~Sitter (dS) for positive, and anti-de~Sitter (AdS) for negative $g_2$.
That momentum tensor thus describes deformations of the dynamical geometry w.r.t.\ (A)dS geometry,
and the parameter $g_1$ has a similar effect as mass has in classical point mechanics.
While it is defined in the denominator of the quadratic momentum term in the Hamiltonian,
it multiplies the conjugate ``velocity'' in the field equation (see below), and also the corresponding quadratic ``kinetic'' term in the gravity Lagrangian.
Greater values of $g_1$ indicate a more ``inert'' space-time with respect to deformation of the curvature tensor versus the (A)dS~geometry, and vice versa.

Furthermore, the canonical equation arising from the variation of Eq.~\eqref{eq:ham-free-grav} with respect to the momentum field $\tilde{k}\indices{_{j}^{\xi\lambda}}$ identifies it with the torsion tensor:
\begin{equation}
 k\indices{_{j}^{\xi\lambda}} = g_3\,S\indices{_{j}^{\xi\lambda}} \, .
\end{equation}
If we assume, for the sake of simplicity, that the Hamiltonian \eqref{eq:ham-free-grav} does not depend on the momentum field $\tilde{k}$,
then the resulting geometry is \emph{torsion free} and \emph{metric compatible}.
The variation principle gives a set of canonical equations of motion, that can be combined to the so called ``consistency equation''~\citep{struckmeier17a, struckmeier21a}, a generalization of Einstein's field equation. Re-written in  coordinate indices and affine connection coefficients, 
it reads:
\begin{align}  \label{def:consistency}
g_1\,\Big( R^{\alpha\beta\gamma\mu}R\indices{_{\alpha\beta\gamma}^{\nu}}
\!\!&-\!\quarter g\indices{^\mu^\nu}R^{\alpha\beta\gamma\xi}  R_{\alpha\beta\gamma\xi} \Big)
+\! {\frac{1}{8\pi G}}\Big(R\indices{^{(\mu\nu)}}\!\!-\!\onehalf
g\indices{^\mu^\nu}\!R
\!-\! {\lambda_0} g\indices{^\mu^\nu} \Big)\!\! \\
&-2g_3\left(S^{\xi\alpha\mu}S\indices{_\xi_\alpha^\nu}-\onehalf 
S^{\mu\alpha\beta}S\indices{^\nu_\alpha_\beta}
-\quarter\,g^{\mu\nu}S_{\xi\alpha\beta}S^{\xi\alpha\beta}\right)=\!\theta^{(\mu
\nu)}.\nonumber
\end{align} 
$g_1$ controls the degree of ``deformation''~\citep{Herranz:2006un} of the Einstein equation vs. General Relativity, and $\theta\indices{^{(\mu\nu)}}$ on the r.h.s.\ of this equation is the symmetric portion of the canonical energy-momentum tensor of matter.\footnote{Notice that the symmetrization arises from the mathematical formalism. In presence of torsion, i.e. if $g_3$ in the last term in the Hamiltonian~\eqref{eq:ham-free-grav} does not vanish, a further trace-free torsion concomitant is added to Eq.~\eqref{def:consistency}. Then the skew-symmetric portions of the Ricci tensor and, if present, the stress-energy tensor, determine the dynamics of torsion. }
\item
A first crucial finding is that, if the torsion is neglected, Eq.~(\ref{def:consistency}) admits both, the de Sitter-Schwarzschild and the de Sitter-Kerr metrics, and is thus compatible with observations on the solar scale. While for pure quadratic gravity an integration constant $\lambda_0$ in the de Sitter-Schwarzschild solution assumes the role of an arbitrary cosmological constant, in quadratic-linear gravity,
\begin{equation} \label{def:constantsg2}
-3g_2 \equiv\lambda_0 \, , 
\end{equation}
is unambiguously fixed~\citep{kehm17}. Moreover, in the non-relativistic limit the classical Newtonian gravitation is recovered if
Newton's gravitational constant $G$ is related to the Einstein terms on the l.h.s.\ of Eq.~(\ref{def:consistency}). This  gives~\citep{struckmeier17a,Vasak:2018gqn} a second relation for the yet unspecified fundamental CCGG constants in the Hamiltonian~\eqref{eq:ham-free-grav} to the empirical constants,
\begin{equation} \label{def:constantsg1}
-2g_1\,g_2 \equiv\frac{1}{8\pi G}\equiv M_{\mathrm{p}}^2 \, .
\end{equation}
$M_{\mathrm{p}}$ is the reduced Planck mass.
Combining these two equations yields
\begin{equation} \label{def:CCgeom}
\lambda_0\equiv\ \frac{3M_{\mathrm{p}}^2}{2g_1} \, .
\end{equation}
It is important to realize that the relation of $\lambda_0$ to $M_{\mathrm{p}}$ emerges from the postulated quadratic-linear ansatz for semi-classical gauge gravity, and is thus a solely classical geometrical contribution to the cosmological constant.
\item  As shown earlier~~\citep{struckmeier17a},  the l.h.s.\ of Eq.~\eqref{def:consistency} is the negative canonical energy-momentum tensor, $-\vartheta^{\,\mu\nu}$, of space-time as derived from the Noether theorem in analogy to that of matter~\citep{struckmeier18a}. Energy and momentum of matter and space-time appear to cancel each other  (``Zero-Energy Universe'', see e.g.~\cite{lorentz1916, levi-civita1917, jordan39,feynman62, Rosen:1994vj, cooperstock95,Hamada:2022fko, Melia:2022ifj}), in analogy to the stress-strain relation in elastic media:
\begin{equation*}
\vartheta^{\,\mu\nu}+\theta^{\,\mu\nu}=0 \, .
\end{equation*}
In absence of matter with even the vacuum energy of matter vanishing, we have
\begin{equation*}
    \theta^{\,\mu\nu} = 0 \, .
\end{equation*}
We expect space-time to sit in its \emph{static} ground state where the momentum fields of gravity vanish. For vanishing momenta of space-time the canonical equations yield
\begin{subequations}
 \begin{align}
 q\indices{_{\,i}^{j\alpha\beta}} &= 0 \implies
 R\indices{^{\,i}_{j\alpha\beta}} = \bar{R}\indices{^{\,i}_{j\alpha\beta}}  \, ; \\
 k\indices{_{\,i}^{\alpha\beta}} &= 0 \implies
 S\indices{_{\,i}^{j\alpha\beta}} = 0 \, ,
\end{align}
\end{subequations}
where $\bar{R}\indices{_{\,i}^{j\alpha\beta}}$ is defined in Eq.~\eqref{def:desitterR}.
By substituting $\bar{R}\indices{_{\,i}^{j\alpha\beta}}$ for ${R}\indices{_{\,i}^{j\alpha\beta}}$ on the l.h.s. of Eq.~\eqref{def:consistency} all terms built from the curvature tensor add up to $6g_1 g_2^2 \,g^{\mu\nu}$, giving indeed \mbox{$\vartheta^{\,\mu\nu}_{\mathrm{vac}} = 0$}. This holds for both, positive and negative values of $g_2$, i.e. for the dS and AdS space-times, respectively.
\item
If, on the other hand, matter exists but is globally in its (quantum) vacuum state, or infinitely far away from any real matter distribution, the stress-energy tensor reduces to
\begin{equation*}
 \theta^{\,\mu\nu} = g^{\mu\nu}\,\theta_{\mathrm{vac}} \, .
\end{equation*}
Now the second crucial point is the observation that space-time is flat (Minkowski) and static in such a configuration, i.e.~$R\indices{_{\,i}^{j\alpha\beta}} = 0$ and $S\indices{_{\,i}^{\alpha\beta}} = 0$, yielding from Eq.~\eqref{def:consistency}
\begin{equation}
-\vartheta^{\,\mu\nu}_{\mathrm{flat}}
= \frac{\lambda_0}{8\pi G} \, g^{\,\mu\nu}
= g^{\mu\nu}\,\theta_{\mathrm{vac}} \, ,
\end{equation}
or with Eq.~\eqref{def:CCgeom}
\begin{equation} \label{eq:lambda0}
-\frac{3M_{\mathrm{p}}^2}{2 g_1} = \frac{\theta_{\mathrm{vac}}}{M_{\mathrm{p}}^2}. 
\end{equation}
This fixes, with the naive estimate $\theta_{\mathrm{vac}} \approx M_{\mathrm{p}}^4$ of the quantum zero-point energy of matter based on a Planck-scale ultra-violet cutoff~\cite{weinberg89},
the value of the coupling constant $g_1 \approx -\nicefrac{3}{2}$.
Since $g_1$ and $g_2$ must be constants and by Eq.~\eqref{def:constantsg1} have the same sign, the ground state of the Universe is the AdS~geometry in this case.
In general we infer that observing the physical geometry of space-time to be Minkowskian in absence of real matter allows to fix the physical value of the constant $g_1$ by the given value of the vacuum energy of matter \citep{Vasak:2018gqn, vasak20}. 
\item
In the following we consider the general situation of a system of real matter embedded in a dynamical space-time with torsion. The total stress-energy tensor in~Eq.~\eqref{def:consistency} consists now of real particular matter and radiation and of the related bulk vacuum energy:
\begin{equation}
\theta^{\mu\nu} \equiv \theta^{\mu\nu}_{\mathrm{tot}} = \theta^{\mu\nu}_{\mathrm{real}} + g^{\mu\nu}\,\theta_{\mathrm{vac}} \, . 
\end{equation}
Then with the relations~\eqref{def:constantsg2} and~\eqref{def:constantsg1} the trace of Eq.~(\ref{def:consistency}) reduces to
\begin{equation} \label{eq:traceCCGG}
	R - \frac{g_3}{M_{\mathrm{p}}^2}\,S^2 + 4\lambda_0 = -\frac{1}{M_{\mathrm{p}}^2} \,\theta_{\mathrm{0}} \, ,
\end{equation}

as the quadratic gravity is trace-free, $\theta_{\mathrm{0}} := g_{\mu\nu}\,T_{\mathrm{0}}^{\mu\nu}$, and $S^2 := S\indices{_\alpha_\beta_\gamma}\,S\indices{^\alpha^\beta^\gamma}$.

Now in this general case we also have to take into account effects of the more complex dynamical geometry and of graviton vacuum fluctuation, giving

\begin{equation} \label{eq:traceCCGG2}
	R  - \frac{g_3}{M_{\mathrm{p}}^2} S^2 \!+ 4\lambda_0  = 
    R_{\mathrm{LC}}   +  R_{\mathrm{geom}}  +  R_{\mathrm{quant}} - \frac{g_3}{M_{\mathrm{p}}^2}\,S^2  \!+  4\lambda_0  =  -\frac{1}{M_{\mathrm{p}}^2} \left( 4 \theta_{\mathrm{vac}}  +  \theta_{\mathrm{real}}
	 \right) \,. 
\end{equation}

Here $R_{\mathrm{LC}}$ is the ``classical'', Levi-Civita version of the Ricci tensor as known from General Relativity, $R_{\mathrm{geom}}$ denotes contribution from dynamical space-time geometry, in the first place from torsion  dynamics which is determined by the canonical equations with the full gravity Hamiltonian including the last term in Eq.~\eqref{eq:ham-free-grav}. $R_{\mathrm{quant}}$ stands for graviton vacuum fluctuations.
This compares directly with the trace of Einstein's field equation with the observed cosmological constant $\Lambda_{\mathrm{obs}}$ 
and the stress-energy tensor $\theta_{\mathrm{real}}^{\mu\nu}$ that is void of vacuum energy:
  \setlength\belowdisplayskip{0pt}
\begin{equation} \label{eq:traceEH}
R_{\mathrm{LC}} + 4\Lambda_{\mathrm{obs}}  = -\frac{1}{M_{\mathrm{p}}^2}  \, \theta_{\mathrm{real}}.
\end{equation}
Combining now Eqs.~\eqref{eq:lambda0}, \eqref{eq:traceCCGG2} and \eqref{eq:traceEH} implies
\begin{equation}
\quarter\left(R_{\mathrm{geom}}+R_{\mathrm{quant}}-8\pi G\,g_3\,S^2\right) 
\equiv \Lambda_{\mathrm{obs}}.
\end{equation}
\end{enumerate}
The concluding statement is thus that in the linear-quadratic gravity ansatz the 
vacuum energy of matter is eliminated, due to the law of Zero-Energy Universe, 
by the  space-time vacuum $\lambda_0$ constructed from the fundamental coupling 
constants of the theory.

The observed cosmological constant $\Lambda_{\mathrm{obs}}$ is merely the current value of a residual correction of geometrical origin based on the dynamics of torsion~\footnote{The impact of the torsion-related corrections of the Einstein equation on cosmology has been discussed in~\cite{minkowski86,Tsamparlis:1981xm,Capozziello:2001mq,Capozziello:2007idz,Poplawski:2010kb,Kranas:2018jdc,vasak20}} with possible contributions from quantum fluctuations of gravitons.
It is a dynamical entity that might be called the \emph{dark-energy function}.
%

\section{Discussion}
The above reasoning shows that the ``cosmological constant problem'' can be explained by an elimination mechanism of the bulk vacuum energies of space-time and matter.
At the first sight that appears as another ``fine tuning'' exercise for $g_1$ of $O(10^{-120})$.
At a second sight it is quite natural, though, to assign values to fundamental constants in order to reproduce empirical findings.
Here fixing the pair $g_1, g_2$ vs. $G, \theta_{\mathrm{vac}}$ goes back to alignment with the weak gravity limit, and to ensuring that static space-time in absence of, or in infinite distance from, any real matter is inertial.

The observed value of the cosmological constant then emerges as a snapshot of a dynamical dark-energy term based on the dynamics of torsion\footnote{
The impact of the torsion-related corrections of the Einstein equation on cosmology has been discussed in \cite{minkowski86, Tsamparlis:1981xm, Capozziello:2001mq, Capozziello:2007idz, Poplawski:2010kb, Kranas:2018jdc, vasak20}. Cosmology with a homogeneous spin density (aka Weyssenhoff fluid) were addressed in Refs. \cite{Obukhov:1987yu, Obukhov:1993pt, boehmer06, Brechet:2007cj, Poplawski:2018ypb, Unger:2018oqo}
A time dependent cosmological constant has been also derived from a naive estimate of vacuum fluctuations~\citep{Santos:2009mu}, string theory~\citep{basilakos19}, and by the renormalization group method~\citep{Myrzakulov:2014hca}.}, and potentially also with contributions from residual quantum fluctuations.

However, uncertainties about the actual value of the constants $g_1$ and $g_2$ remain due to the yet unknown value of the vacuum energy of matter.
Albeit the naive estimate of $\theta_{\mathrm{vac}} \approx M_{\mathrm{p}}^4$ is at the heart of the cosmological constant problem, it is not supported by all field- or string-theoretical calculations, and can vary in value and sign almost arbitrarily.
Assuming for example the extreme case supported by the string theory, $\theta_{\mathrm{vac}} \to 0$, then according to Eq.~\eqref{eq:lambda0} we find~\citep{Vasak:2018gqn}
$g_1 \to \infty$.
A vacuum energy density of matter in the meV range, a value discussed previously for example in Ref.~\cite{Prat:2021xlz}, leads 
to $g_1 \sim -10^{119}$.

\section{Conclusion}
The motivation for this paper was to shed new light on the cosmological constant problem derived originally by Weinberg~\citep{weinberg89} from the naive assessment of the vacuum energy of matter, $\theta_{\mathrm{vac}} \sim M_{\mathrm{p}}^4$, based on the Planck energy cutoff.

The facts collected here stand to reason that the combination of linear (Einstein-Cartan) and (trace-free) quadratic gravity in the classical gauge-field theory of gravity, and the empirical knowledge of Newton's constant and  astronomical observations, can explain both the existence \emph{and} to a significant extent also the magnitude of the cosmological constant. In fact, the above reasoning is valid regardless of the actual value of $\theta_{\mathrm{vac}}$, for a compensating value of $g_1$ can always be found as the solution of Eq.~\eqref{eq:lambda0}.

This is accomplished by applying a rigorous mathematical framework in a consistently designed gauge field theory of gravity to derive a geometrical term eliminating the bulk of quantum vacuum energy of matter. 
The residual, physical cosmological constant turns out to be just a snapshot of an underlying dynamical entity, a dark-energy function built from torsion of space-time and quantum fluctuations of gravitons.

However, ambiguities remain in the value of the vacuum energy of matter and the 
exact form of the geometrical and matter related vacuum corrections.
Designing independent measurements of  $g_1$ and analyzing a wider variety of dynamical torsion models furnishing the dynamics of the cosmological dark energy will thus be the next key steps to resolving those remaining ambiguities.

Work along these lines is in progress. 

\section*{Acknowledgements}
The authors are indebted to the ``Walter Greiner-Gesell\-schaft zur F\"{o}rderung
der physi\-ka\-lischen Grundlagenforschung e.V.'' (WGG) in Frankfurt for their support.\,
H.St.\ acknowledges the Judah M. Eisenberg Laureatus Professur of the WGG at the Fachbereich Physik at
Goethe Universit\"at Frankfurt.\, DV and JK especially thank the Fueck Stiftung for support. \\
The authors also wish to thank Armin van de Venn, David Benisty, Stefan Hofmann, Eduardo Guendelman and Peter Hess for valuable discussions.

\pagebreak
\section*{References}
\input{Vasak_ARXIV.bbl}
\end{document}
%

%% file: Vasak_ARXIV.bbl
%